\documentclass[aps,prl,twocolumn,amsmath,amssymb,showpacs]{revtex4}

\usepackage{graphicx,calc,psfrag}
\usepackage{bm}

\begin{document}

\title{Diffraction-contrast imaging of cold atoms}
\newcommand{\unimelb}{School of Physics, University of Melbourne,
  Victoria 3010, Australia}
\newcommand{\tue}{Faculty of Applied Physics, Eindhoven University
  of Technology, The Netherlands}

\author{L.~D.~Turner}
\email{l.turner@aip.org.au}
\author{K.~F.~E.~M.~Domen}
\altaffiliation{Present address: \tue}
\author{R.~E.~Scholten}
\affiliation{\unimelb}

\date{\today}

\begin{abstract}
  We consider the inverse problem of in-line holography, applied to
  minimally-destructive imaging of cold atom clouds.  Absorption
  imaging near resonance provides a simple, but destructive
  measurement of atom column density. Imaging off resonance greatly
  reduces heating, and sequential images may be taken. Under the
  conditions required for off-resonant imaging, the
  generally-intractable inverse problem may be linearized. A
  minimally-destructive, quantitative and high-resolution image of the
  atom cloud column density is then retrieved from a single
  diffraction pattern.
\end{abstract}

\maketitle

\newcommand{\f}{\mathcal{F}\!}
\newcommand{\rvec}{\mathbf{r}}
\newcommand{\xvec}{\mathbf{x}}
\newcommand{\uvec}{\mathbf{u}}
\newcommand{\avec}{\mathbf{a}}
\newcommand{\intall}{\int_{-\infty}^{+\infty}}
\newcommand{\er}[1]{Eq.\,(\ref{#1})}
\newcommand{\fti}{\f\left[I(\xvec,z)\right]}
\newcommand{\rplus}{\xvec\!+\!\lambda z \uvec/2}
\newcommand{\rminus}{\xvec\!-\!\lambda z \uvec/2}
\newcommand{\micron}{\mathrm{\mu m}}
\newcommand{\dform}{\,\mathrm{d}}
\newcommand{\clcdot}{\!\cdot\!}
\newcommand{\mm}{\,\mathrm{mm}}
\newcommand{\zeff}{z_\text{eff}}
\newcommand{\axisstyle}{\footnotesize}
\newcommand{\mum}{\mu \text{m}}
\newcommand{\thuz}{\tilde{h}(u;z)}
\newcommand{\plzu}{\pi\lambda z u^2}

\begin{figure}[b]
  \centering
  \includegraphics{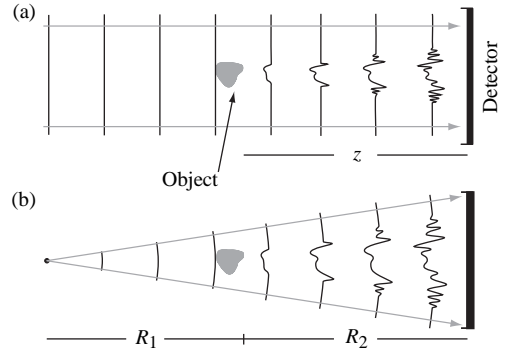}
  \caption{Recording a diffraction pattern. Vertical lines represent
    intensity profiles. In part (b), point-source illumination
    magnifies the diffraction pattern.\label{f-pointproj} }
\end{figure}

The simplest optical measurement of object structure is made by
illuminating the object with radiation and recording the diffraction
pattern produced (Fig.~\ref{f-pointproj}). The inverse problem of
retrieving the structure of a non-crystalline object from its Fresnel
diffraction pattern has been studied since Gabor's first incomplete
solution, known as in-line holography~\cite{gabor}.  This
Communication solves the particular inverse problem of retrieving the
structure of a cold atom cloud from a single diffraction pattern.

Minimally-destructive imaging of cold atoms requires weak absorption.
High-resolution imaging also places constraints on phase shifts. Under
these assumptions, we derive a linear solution which retrieves the
column density of the atom cloud from a single diffraction pattern. We
apply this solution to demonstrate off-resonant imaging of a
cold atom cloud without beamsplitters, phase-plates, or imaging
optics.

Gabor's in-line holography recovers an approximation of the original
wavefield by illuminating a photographic transparency of the
diffraction pattern. The reconstructed wavefield is contaminated by
the superimposition of an out-of-focus \emph{twin
  image}~\cite{twinimage}.  Other forms of holography use a reference
beam to record an interference pattern, rather than a diffraction
pattern, and so separate the twin image~\cite{leith}.  In this
Communication we demonstrate a \emph{non-holographic} method of
retrieval. Such methods have been proposed when it is inconvenient or
impossible to generate a coherent reference beam. The first step is
common with Gabor's method: a diffraction pattern is recorded without
the need for optics such as lenses and beamsplitters.  In the second
step, rather than reconstructing the wavefield by optical propagation
(physical or numerical), an image is extracted by solving an inverse
problem with specified constraints.

There is insufficient information in a single intensity image to
retrieve both the amplitude and the phase of the wavefield.  This
information deficit may be balanced, and the inverse problem solved,
if the object is assumed to be purely absorbing~\cite{onural} or
purely phase-shifting~\cite{cloetens}, but these assumptions are
seldom valid in practice. 

Instead, we present a single-image solution based on the assumption of
a monomorphous object (one made of a single material), so that both
the phase-shift $\phi$ and the absorption $\mu$ of the object are
proportional to the column density of material along the optical path
$\rho(\xvec)=\int_{-\infty}^{0}N(\rvec)\dform z$:
\begin{equation}
  \label{e-homog}
  \mu(\xvec) = k\beta \rho(\xvec) \quad \text{and} \quad
  \phi(\xvec) = k\delta \rho(\xvec). 
\end{equation}
The variable $\xvec$ represents coordinates in a plane transverse to the
incident wave propagating along the $z$-axis and $k=2\pi/\lambda$ is
the wavenumber for illuminating radiation of wavelength $\lambda$. 
The absorption and phase coefficients $\beta$ and $\delta$ correspond
to a refractive index of the form 
\begin{equation}
  \label{e-monorefind}
  n=1+N(\rvec)(\delta+i\beta),  
\end{equation}
with $N(\rvec)$ the atom number density. This mono\-morphous object
assumption has been used in compensating defocus and
spherical aberration in electron micrographs~\cite{erickson} and
transport-of-intensity imaging~\cite{paganin}. 

Immediately after an optically thin object (Fig.~\ref{f-pointproj}(a)), an incident scalar
plane-wave of amplitude $f_0$ becomes
$f(\xvec)=f_0\exp\big(-\mu(\xvec)+i\phi(\xvec)\big)$, and this
wavefield may be propagated through a distance $z$ by the Fresnel
transform
\begin{equation}
  \label{fresnel} 
  f(\xvec,z) = \frac{\exp(ikz)}{i\lambda z} \intall f(\xvec)
          \exp\left(\frac{i \pi}{\lambda z}
          |\xvec-\xvec'|^2\right)\dform\xvec'. 
\end{equation}
The Fresnel approximation agrees closely with the complete scalar
diffraction theory, except for propagation at large angles to the axis
or within a few wavelengths of the object. Optical detectors measure
intensity $I=|f|^2$ and it can be shown~\cite{guigay77} that the
Fourier transform $\f$ of the diffracted intensity measured at $z$ can
be expressed in terms of the object-plane wavefield $f(\xvec)$ as
\begin{align}
  \label{guigay} 
  \fti = \intall & f^{\ast}(\rplus)f(\rminus)\nonumber\\
  \times & \exp(-2\pi i \xvec \cdot \uvec)\dform\xvec,
\end{align}
in which $\uvec$ is the spatial frequency conjugate to
$\xvec$. Written in terms of absorption and phase-shifts, this is
\begin{align}
  \label{e-ctfstep1}
  \fti = I_0 \intall &
  \exp\{-\mu(\rplus)-\mu(\rminus)\nonumber\\
   + & i\left[\phi(\rminus)-\phi(\rplus)\right]\}\nonumber\\
  \times & \exp(-2\pi i  \xvec\cdot\uvec)\dform\xvec. 
\end{align}
Assuming both real and imaginary parts of the exponential are small,
we expand and apply the Fourier shift theorem to yield:
\begin{align}
  \label{e-weakctf}
  \fti=I_0 \big(\delta(\uvec) - & 2\cos (\plzu)
  \f[\mu(\xvec)] \nonumber\\ 
  + & 2\sin(\plzu) \f[\phi(\xvec)]\big). 
\end{align}
This expression~\cite{menzel,turner} relates absorption and
phase-shift to the intensity of the diffraction pattern. The
linearizing assumption used in obtaining~\er{e-weakctf} implies:
\begin{align}
  \label{abscond}
  & 2\mu(\xvec) \ll 1 \\ 
  \label{phasecond}
  \text{and}\qquad & \left|\phi(\rplus)-\phi(\rminus)\right| \ll 1. 
\end{align}
The object must not be strongly absorbing, but it need not be
completely transparent.  The phase-shift should obey the finite
difference condition~\er{phasecond}, which restricts large variations
in the phase-shift to coarse structures in the object. Note that weak
phase-shift ($|\phi(\xvec)|\ll 1$) is sufficient to
satisfy~\er{phasecond} but is \emph{not} necessary~\cite{turner}. This
phase condition may always be met at small $z$, but phase objects of
many radians thickness may require impractically small propagation
distances, and phase shifts of order $1$\,radian are preferable. 

\begin{figure}[t]
  \centering
  \includegraphics{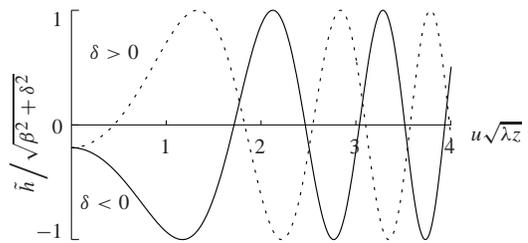}
  \caption{The contrast transfer function $\tilde{h}$, normalized,
    for phase-advancing (solid line) and
    phase-retarding (dashed line) monomorphous objects. At low
    spatial frequencies the contrast approaches the value for an
    in-focus absorption image.  
    \label{f-ctf} }
\end{figure}

For monomorphous objects obeying~\er{e-monorefind}, there is then a
linear shift-invariant relation between the normalized contrast
$I/I_0-1$ and the column density $\rho$:
\begin{equation}
  \label{e-monoctf}
  \f\left[\frac{I-I_0}{I_0}\right]=2k\left(\delta\sin(\pi\lambda z
    u^2)-\beta\cos (\pi \lambda z u^2)\right)\f[\rho(\xvec)]. 
\end{equation}
The factor $\thuz=\delta\sin(\plzu)-\beta\cos(\plzu)$ is termed the
\emph{contrast transfer function} (CTF), and is plotted in
Fig.~\ref{f-ctf} for positive and negative values of the phase
coefficient $\delta$.  Equation~\ref{e-monoctf} can be solved formally
for $\rho$, but the zeros in the CTF render the retrieval an
\emph{ill-posed} inverse problem. 

The inverse problem may be \emph{regularized}, for example by the
Tikhonov method~\cite{tikhonov}. Rather than dividing $\f[I/I_0-1]$
by $\thuz$, the Tikhonov filter retrieves the column density by the
modified division
\begin{equation}
  \label{e-tikhonov}
  \rho(\xvec) =
  \frac{1}{2k}\,\f^{-1}\!\left[\frac{\thuz}{\tilde{h}^2(u;z)+\alpha^2} 
    \f\left[\frac{I-I_0}{I_0}\right]\right]
\end{equation}
which closely approximates division by the CTF except at spatial
frequencies where the CTF is near zero.  Larger values of the Tikhonov
parameter $\alpha$ reduce the amplification of noise in the retrieval
process, but at the expense of image distortions. Smaller values yield
less distorted but noisier retrievals; a normalized alpha value
of $0.2$ was used. Algorithmic optimization of $\alpha$ is
possible, for example using Fourier-wavelet regularized
deconvolution~\cite{forward}. 

It is clear from Fig.~\ref{f-ctf} that the solution is more stable if
the object advances the phase of the incident wave ($\delta<0$), and
the CTF zero-crossing at low spatial frequencies is avoided.  The
column density may be retrieved for phase-retarding objects
($\delta>0$) but, as shown in Fig.~\ref{f-ctf}, the focusing action of
the phase-shift cancels the absorption contrast at low spatial
frequencies and lower quality retrievals result. 

If the diffraction pattern is re-imaged by a lens, the system may be
defocused behind the object so that the effective propagation distance
$z$ is negative. It follows from~\er{e-monoctf} that the
sign-condition on $\delta$ is then reversed. For lensless imaging,
negative $z$ cannot be achieved and the object should be
phase-advancing. 

Magnified images can be retrieved even without lenses.  Rather than
using plane-wave illumination, a point-source of light a distance
$R_1$ before the object produces a spherical wave incident on the
object (Fig.~\ref{f-pointproj}(b)).  The diffraction at detector
distance $R_2$ is magnified by the geometric factor $M=(R_1+R_2)/R_1$,
but is otherwise identical to the plane-wave pattern of
Fig.~\ref{f-pointproj}(a) at the effective
propagation distance $\zeff=R_2/M$~\cite{pogany}.  

Conventional optical materials are phase-retarding ($\delta>0$) but
for x-ray imaging~\cite{turner}, and for imaging atomic gases with
light blue-detuned from an atomic resonance, the phase is advanced. 
We now show that our solution to the diffraction imaging
inverse problem is exactly suited to off-resonant imaging of cold atom
clouds. 

To date, all measurements of ground-state BECs have been made with
near-resonant optical probes. On-resonance absorption imaging is
destructive for most BEC configurations. Imaging with an off-resonant
probe reduces heating due to spontaneous emission, with the cloud
instead shifting the phase of the probe beam. Dark-field~\cite{dark}
and Zernike phase-contrast~\cite{zernike} techniques of phase
microscopy have been used to render these phase-shifts visible and
hence obtain sequential, minimally-destructive images of BEC. Other
methods investigated for off-resonant imaging include an
interferometric technique equivalent to off-axis image
holography~\cite{kadlecek} and a propagation-contrast method based on
transport-of-intensity~\cite{turnerweber}. Although
minimally-destructive imaging has been crucial to observing many
dynamic processes in BEC, destructive absorption imaging is still
widely used. 

As shown above, free space propagation produces phase contrast without
optics. Gaussian fits to atom cloud images (peak column density and
diameters only) have been extracted from fitting diffraction
patterns~\cite{meschede}. Our solution to the inverse problem
retrieves detailed column-density \emph{images} of cold atom clouds,
without requiring Zernike phase-plates or interferometry. 

In the two-level approximation, the refractive index of an atomic gas 
is 
\begin{equation}
  \label{refind}
  n = 1+ N(\rvec)\frac{\sigma_0\lambda}{4\pi}\frac{i-2\Delta}{1+4\Delta^2}
\end{equation}
where $N$ is the number density of atoms, $\Delta$ is the detuning in
natural linewidths and $\sigma_0$ is the resonant cross-section
($3\lambda^2/2\pi$ for closed transitions). Comparison
with~\er{e-monorefind} confirms that such an atomic cloud is a
monomorphous object, with absorption and phase-shifts through the
cloud given by~\er{e-homog}. Provided that the atom cloud satisfies
the weak-absorption condition~\er{abscond} and limited-phase
condition~\er{phasecond}, the CTF relation~\er{e-monoctf} applies. 
Substituting the $\beta$ and $\delta$ coefficients from~\er{refind}
yields the cold atom CTF
\begin{equation}
  \label{ctfatom}
    \thuz = -\frac{\sigma_0}{2k(1+4\Delta^2)}\big[
    2\Delta \sin(\plzu)- \cos (\plzu)\big],
\end{equation}
which is then used in~\er{e-tikhonov} to retrieve the column density
of the atom cloud.  If the detuning is blue of resonance, the atom
cloud advances the phase of the incident light, and the low-frequency
CTF zero is avoided. 

An important feature of this application to cold atom imaging is the
regularizing effect of residual absorption.  At zero spatial
frequency, the CTF falls to $\sim 1/2\Delta$ of its maximum value, but
does not vanish completely as it would for a pure phase object.  Even
small residual absorption is sufficient to stabilize the retrieval and
then the Tikhonov modified form~\er{e-tikhonov} need only be used at
higher spatial frequencies above $u_\text{min}=1/\sqrt{2\lambda z}$. 
Such partial regularization greatly reduces distortion, and retrievals
approach the optimal linear estimate (Wiener filter) which can only be
calculated with full advance knowledge of the object power spectrum. 

This linearization of the inverse problem is only valid if the atom
cloud meets the absorption and phase
conditions~(\ref{abscond},\ref{phasecond}). Minimally-destructive
imaging necessarily obeys the weak-absorption condition. The phase
condition~\er{phasecond} also broadly concurs with physical
constraints due to refraction and resolution. Light refracted by the
object must remain within the numerical aperture of the lens (or
detector)~\cite{dark}.  It follows that structures at the diffraction
limit of the imaging system should have phase variations less than one
radian. Objects satisfying this `thin object' condition -- that the
detailed structure of the object must vary by less than a radian --
are likely to also satisfy the slowly-varying phase condition~\er{phasecond}. 

Detuning the probe light by $\rho_\text{max}\sigma_0/4$ full
linewidths from resonance reduces the peak phase-shift to order one
radian.  BECs typically have resonant optical-densities
$\rho_\text{max}\sigma_0\approx 300$ and so detunings must be of order
$100\Gamma$ to meet the refraction condition. At such detunings, many
images may be taken before the cloud is appreciably heated. In the
shot-noise limit, further increasing detuning and intensity does not
improve the SNR beyond a limiting value, and in the presence of
technical noise will reduce the SNR. It has been shown that this
SNR limit is determined only by the number of spontaneous
emission events and condensate parameters~\cite{close}. 

\begin{figure}[tb]
 \includegraphics{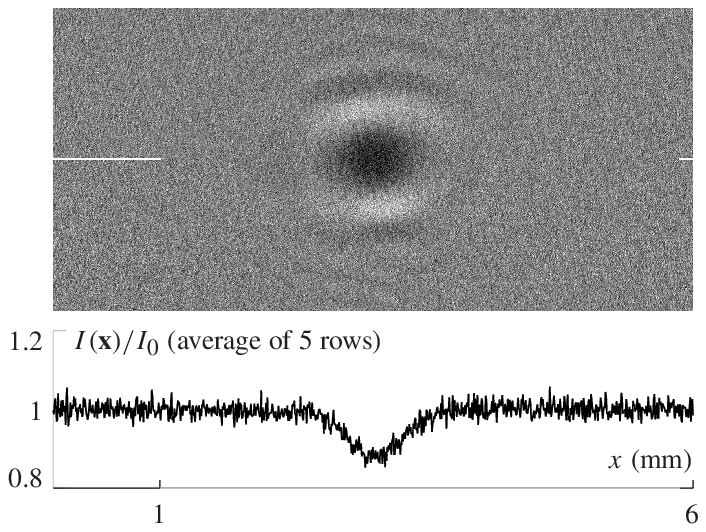}\\
 \vspace*{2mm}
 \includegraphics{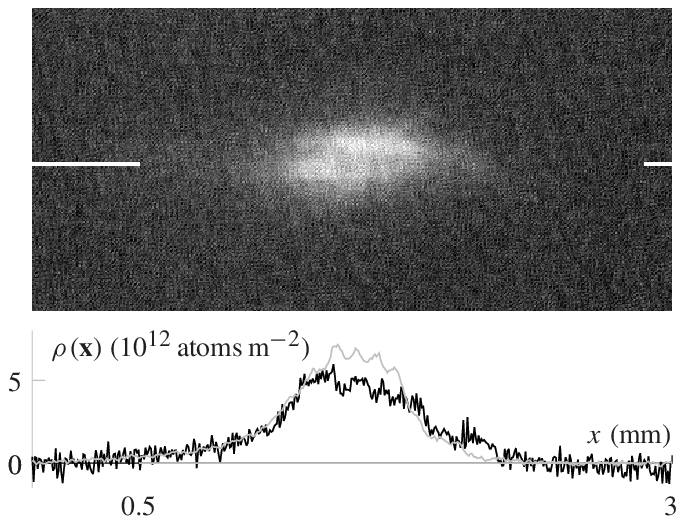}
 \caption{Above, the recorded diffraction pattern. Below, the
   column density retrieved using~\er{e-tikhonov}, shown at twice the
   magnification of the diffraction pattern. Plots averaged over the
   central five rows of pixels are shown below the images. The
   overplotted gray trace shows the column density calculated from an
   on-resonance in-focus absorption image. 
    \label{f-diffrac}}
\end{figure}

In proof-of-principle experiments, the point-projection configuration
of Fig.~\ref{f-pointproj}(b) was used to image a cold atom cloud of
peak resonant optical density 2.2. A weak linearly-polarized probe
beam, detuned $+3.1\Gamma$ from the $^{85}$Rb 4S$_{1/2}(F\!=\!3)$
$\rightarrow$ 5P$_{3/2}(F\!=\!4)$ transition, diverged from the
cleaved endface of a single-mode optical fiber, expanding for
$R_1=125$\,mm before passing though a cloud of cold atoms around
$300\,\micron$ in diameter held in a magneto-optical trap (MOT) vacuum
chamber. Trapping beams were turned off for the $300\,\mu s$ duration
of the probe pulse. The beam propagated a further distance
$R_2=155$\,mm to a CCD camera (Photometrics Coolsnap HQ,
1392$\times$1040 pixels, $6.45\,\micron$ pitch), producing the
diffraction pattern shown in Fig.~\ref{f-diffrac}.  The column-density
image was retrieved using~\er{e-tikhonov}; retrievals take around one
second on a Pentium-III processor using standard discrete Fourier
transform algorithms~\cite{software}.

It is usually not possible to measure the propagation distances $R_1$
and $R_2$ accurately enough to produce an optimal retrieval. Instead,
the retrieval is performed with the contrast transfer function $\thuz$
evaluated at various values of $z$ until a sharp image is retrieved. 
Thus focusing is performed in software when retrieving, rather than by
adjusting lens positions when imaging.  As a corollary, the retrievals
show holographic depth-of-field: one diffraction pattern can be used
to retrieve images at many different $z$ values. The very real
problems of focusing the optics, and of the limited depth-of-field
inherent in high-resolution imaging, are completely obviated. 

While the precise setting and knowledge of the propagation distance is
immaterial, its coarse setting affects the shape and contrast of the
diffraction pattern and hence the signal-to-noise ratio (SNR) of the
retrieved image. At short distances only residual absorption contrast
will be rendered. At large propagation distances and for small
phase-shifts, the root-mean-square SNR approaches 71\% of that
obtained with the Zernike technique~\cite{turnerphd}. 

A further advantage of the point-projection configuration is the
absence of lenses, and their resolution-limiting aberrations. 
Diffraction contrast may also be produced by defocusing an existing
absorption imaging system, which may be more convenient than placing
the camera very close to the object. The advantages of holographic
depth-of-field and post-hoc focusing are retained. 

In practice, the propagation distance is constrained by resolution
limits. For a detector of diameter $D$, the minimum resolvable line
spacing in the retrieved image is of order $2\lambda R_2/D$, as it is
for a lens of the same diameter in the same position.  The $R_1$
distance should then be chosen to provide sufficient magnification
that resolution is limited by diffraction and not by the pixel size. 
In our experiment, optical access limited the resolution to
$30\,\micron$. We can predict that a BEC in a glass cell imaged with
$R_1=12$\,mm and $R_2=60$\,mm on a $D=25$\,mm CCD with $9\,\micron$
pixels yields a pixel resolution of $3\,\micron$ and a
diffraction-limited resolution of $3.7\,\micron$. Further, the CTF
depends on $z$, $\lambda$, $\Delta$ and $u$. Once $z$ is set by
`focusing', the remaining quantities are readily measured to better
than $1\%$.  Such well-defined parameters and the lack of lens
aberrations yield highly quantitative column density measurements. 

We solve the inverse problem of retrieving a quantitative column
density image from a single diffraction pattern by exploiting the
proportionality between absorption and phase shift through a
single-material object. The predicates of the solution are uniquely
suited to imaging cold atom clouds. Lens aberrations are precluded
by avoiding the need for image formation entirely. Beam-splitters,
phase-plates and other optical elements are also unnecessary. The
holographic record allows refocusing after the image has been
acquired. The solution can be used with existing absorption imaging
systems simply by defocusing the imaging lens.  We calculate
near-wavelength resolution when using point-projection to image
Bose-Einstein condensates.

\end{document}